\documentclass{article}
\usepackage{epsfig}

\begin{document}

\def\be{\begin{equation}}
\def\ee{\end{equation}}
\def\ba{\begin{eqnarray}}
\def\ea{\end{eqnarray}}

\title{Crossing of two Coulomb--Blockade Resonances\\}

\author{Hans A. Weidenm\"uller\\
Max-Planck-Institut f\"ur Kernphysik,
D-69029 Heidelberg, \\
Germany}

\date{\today}

\maketitle

\begin{abstract}

We investigate theoretically the transport of non--interacting
electrons through an Aharanov--Bohm (AB) interferometer with two
quantum dots (QD) embedded into its arms. In the Coulomb--blockade
regime, transport through each QD proceeds via a single resonance. The
resonances are coupled through the arms of the AB device but may also
be coupled directly. In the framework of the Landauer--B\"uttiker
approach, we present expressions for the scattering matrix which
depend explicitly on the energies of the two resonances and on the AB
phase. We pay particular attention to the crossing of the two
resonances.


\end{abstract}

\section{Introduction}\label{int}

The crossing of two Coulomb--blockade resonances was studied in two
recent experiments~\cite{enss,huet}. In both cases, two quantum dots
(QD) were imbedded into the arms of an Aharanov--Bohm (AB)
interferometer. By changing the parameters of the experiment (various
gate voltages and the magnetic flux through the AB device), it was
possible to study the crossing properties of two isolated
Coulomb--blockade resonances, one each due to one of the two QDs. In
the present paper we present a theoretical framework for the analysis
of both experiments.

Figure~1 shows a schematic representation of both experiments. The AB
ring contains the two QDs labelled QDL and QDR where L and R stand for
left and right, respectively. The QDs are separated by barriers from
the rest of the AB device. The latter consists of two parts. In
Figure~1, the lower (upper) part is labelled 1 (2, respectively). Both
parts are coupled to the outside world by a number of leads. In
Figure~1, this number is two (three) for part 1 (part 2, respectively).
In our theoretical treatment, the number of leads coupled to each part
will be arbitrary. Typically, one of the leads coupled to part 1 (part
2) serves as source (sink, respectively) for the electrons. While the
two QDs are not coupled directly to each other in the first
experiment~\cite{enss}, such a coupling does exist in the second
experiment~\cite{huet}. This coupling is indicated schematically by the
dotted horizontal line representing the wire connecting QDL and QDR. In
Ref.~\cite{huet}, the strength of that coupling was controlled by a
further gate. Figure~1 does not show the plunger gates which make it
possible to control the energies of the Coulomb blockade resonances in
either QD. Thereby it is possible to have the energies of both Coulomb
blockade resonances coincide. Experimentally, such crossings are seen
in three--dimensional plots of the conductance versus the plunger gate
voltages $V_{\rm L}$ and $V_{\rm R}$ applied on QDL and QDR,
respectively. Each Coulomb--blockade resonance corresponds to a ridge.
The ridges of resonances in QDL (QDR) run essentially parallel to
$V_{\rm R}$ ($V_{\rm L}$, respectively). The crossing of two such ridges
marks the crossing of two Coulomb--blockade resonances. The coincidence
of two resonances also affects the interference pattern of the
transmission of an electron through the AB device. This pattern depends
upon the magnetic flux $\Phi$ through the device. The flux is due to a
homogeneous magnetic field perpendicular to the plane of the drawing.
We are interested in weak magnetic fields only. (We recall that for a
complete AB oscillation, the magnetic field strength typically changes
at most by several ten mT.) Therefore, we take into account only the AB
phase due to the magnetic flux and neglect the influence of the magnetic
field on the orbital motion of the electron. Gauge invariance then
allows us to link the AB phase to the passage of the electron through a
particular part of the AB device. In the absence of a direct coupling
between the two QDs (i.e., without the dotted line in Figure~1), we
choose the barrier separating QDL from part 1. Whenever the electron
leaves (enters) QDL for part 1 (from part 1), it picks up the phase
factor $\exp( 2 i \pi \Phi / \Phi_0)$ ($\exp( - 2 i \pi \Phi / \Phi_0)$,
respectively) where $\Phi_0$ is the elementary flux quantum. For brevity,
we write the phase factor as $\exp ( i \phi)$. In the presence of a
direct link between the two QDs, the topology of the AB interferometer
changes from that of a ring to that of a figure eight, and we use a
different convention in Section~\ref{link}.

In Section~\ref{hami} we define the Hamiltonian for the system. In
Section~\ref{scat} we use the Landauer--B\"uttiker approach and present
the generic form of the scattering matrix which describes the
experimental setup of Ref.~\cite{enss}. In Section~\ref{cross}, this
scattering matrix is analysed especially with regard to the crossing of
two Coulomb--blockade resonances. In Section~\ref{link}, we generalize
our treatment to include the setup of Ref.~\cite{huet}. In
Section~\ref{summ} we list the approximations and summarize our
approach and results. Moreover, we address some of the approximations
made. In particular, we discuss the neglect of the mutual Coulomb
interaction between the two electrons which are added to the system as
the resonances become populated, and that of the Coulomb interaction
between each of these electrons and those on the dots. We also address
the role of the spins of both quantum dots and of the two added
electrons. Throughout the paper, we disregard temperature averaging for
simplicity. Likewise, we disregard decoherence effects although these
are known to play some role in the actual experiments. We do so because
part of the transport through the device is known to proceed coherently.
Only this part will display a dependence on the AB phase. Moreover,
decoherence has been thoroughly discussed in the literature, see, for
instance, Ref.~\cite{burk}.

\section{Hamiltonian}
\label{hami}

In defining the Hamiltonian of the system, we proceed in full analogy
to Refs.~\cite{hack,weid}. These papers addressed the AB phase for a
single QD placed in one of the arms of an AB interferometer. (For a
review of work on this problem, see Ref.~\cite{hack1}). We introduce
fictitious barriers separating parts 1 and 2 of the AB device from
the attached leads. Likewise, we consider parts 1 and 2 as separated
from the two QDs. We impose boundary conditions on all these barriers
such that as a result, we obtain self--adjoint single--particle
Hamiltonians $H_{\rm lead}$ for the leads, $H_1$ and $H_2$ for the now
separated parts 1 and 2, and $H_{\rm L}$ and $H_{\rm R}$ for the two
QDs labelled QDL and QDR, respectively. Here $H_{\rm lead}$ possesses
a continuous spectrum while the spectra of $H_1$, $H_2$, $H_{\rm L}$
and $H_{\rm R}$ are discrete. We label the leads attached to part 1
(part 2) by $s = 1, \ldots, S$ (by $t = 1, \ldots, T$, respectively).
The transverse modes (channels) in lead $s$ ($t$) are labelled $a = 1,
\ldots, N_s$ ($a = 1, \ldots, N_t$, respectively), and correspondingly
for the creation and annihilation operators $c^{\dagger}$ and $c$. The
associated energies are labelled $\epsilon$. The eigenvalues of $H_1$
($H_2$) are labelled $E_{1 j}$ ($E_{2 j}$), with $j = 1, \ldots,
\infty$ and associated creation and annihilation operators
$c^{\dagger}_{1 j}$ ($c^{\dagger}_{2 j}$), and $c_{1 j}$ ($c_{2 j}$,
respectively). We assume that transport through either QD occurs in
the Coulomb--blockade regime where the intrinsic widths of individual
resonances are small compared to their spacings. (The spacing
includes, of course, the charging energy). We also assume that the
temperature is small in comparison with the spacings. Under these
conditions, it is legitimate to assume that transport through either
QD is dominated by a single Coulomb--blockade resonance. We believe
that this situation is met or nearly met in the experiments of
Refs.~\cite{enss, huet}. Thus, we admit only a single bound state with
energy $E_{\rm L}$ ($E_{\rm L}$) in QDL (QDR), with associated
creation and annihilation operators $d^{\dagger}_{\rm L}$
($d^{\dagger}_{\rm R}$) and $d_{\rm L}$ ($d_{\rm R}$, respectively).
The energies $E_{\rm L}$ and $E_{\rm L}$ include the charging
energies. Altogether, we have
\ba
H_{\rm lead} &=& \sum_{s a} \int {\rm d} \epsilon \ \epsilon \
c^{\dagger}_{s a}(\epsilon) c_{s a}(\epsilon) + \sum_{t a} \int {\rm
  d} \epsilon \ \epsilon \ c^{\dagger}_{t a}(\epsilon) c_{t
  a}(\epsilon) \ , \nonumber \\
H_1 &=& \sum_j E_{1 j} c^{\dagger}_{1 j} c_{1 j} \ , \nonumber \\
H_2 &=& \sum_j E_{2 j} c^{\dagger}_{2 j} c_{2 j} \ , \nonumber \\
H_{\rm L} &=& E_{\rm L} d^{\dagger}_{\rm L} d_{\rm L} \ , \nonumber \\
H_{\rm R} &=& E_{\rm R} d^{\dagger}_{\rm R} d_{\rm R} \ .
\label{01}
\ea
Hopping between the separate parts is induced by interaction terms
containing tunneling matrix elements,
\ba
H_{{\rm lead} 1} &=& \sum_{s a; j} \int {\rm d} \epsilon (V_{s a; 1
  j}(\epsilon) c^{\dagger}_{s a}(\epsilon) c_{1 j} + {\rm h.c.}) \ ,
\nonumber \\ 
H_{{\rm lead} 2} &=& \sum_{t a; j}  \int {\rm d} \epsilon (V_{t a; 2
  j}(\epsilon) c^{\dagger}_{t a}(\epsilon) c_{2 j} + {\rm h.c.}) \ ,
\nonumber \\
H_{1 {\rm L}} &=& \sum_{j} (V_{1 j; L} c^{\dagger}_{1 j} d_{\rm L} +
{\rm h.c.}) \ , \nonumber \\
H_{2 {\rm L}} &=& \sum_{j} (V_{2 j; L} c^{\dagger}_{2 j} d_{\rm L} +
{\rm h.c.}) \ , \nonumber \\
H_{1 {\rm R}} &=& \sum_{j} (V_{1 j; R} c^{\dagger}_{1 j} d_{\rm R} +
{\rm h.c.}) \ , \nonumber \\
H_{2 {\rm R}} &=& \sum_{j} (V_{2 j; R} c^{\dagger}_{2 j} d_{\rm R} +
{\rm h.c.}) \ .
\label{02}
\ea
The direct coupling of QDL and QDR (dashed line in Figure~1) is given
by
\be
H_{\rm L R} = V_{\rm L R} (d^{\dagger}_{\rm L} d_{\rm R} + {\rm h.c.})
\ .
\label{03}
\ee 
In the absence of any direct coupling between QDL and QDR ($V_{\rm L
  R} = 0$) we use gauge invariance to put the entire AB phase onto a
single one of the barriers. Without loss of generality we choose the
barrier separating QDL and part 1. Then, all the matrices $V$ in
Eqs.~(\ref{02}) are real and symmetric except for $V_{1 j; L}$ which
obeys
\be
V_{1 j; L} \exp(- i \phi) = V_{L; 1 j} \exp(i \phi) = v_{1 j; L} 
\label{04}
\ee 
with $v_{1 j; L}$ real and symmetric. For $V_{\rm L R} \neq 0$, a
modification is neccessary and discussed in Section~\ref{link} below.
The Hamiltonian $H$ of the system is the sum of the terms defined by
Eqs.~(\ref{01}) to (\ref{03}). We have not considered the possibility
of spin--orbit coupling on either QD.

We have been very explicit in the construction of $H$. The reason is
that we wanted to show that $H$ is a sum of single--particle
Hamiltonians. This fact allows us to use the Landauer--B\"uttiker
approach to describe transport through the system. The ensuing use of
the scattering matrix enables us to display explicitly the phase-- and
energy--dependence of the conductance coefficients. We have omitted
the spins of as well as any possible interaction between the two
electrons which will eventually populate the two resonances caused by
$E_{\rm L}$ and $E_{\rm R}$. These points are taken up in
Section~\ref{summ}.

\section{Scattering Matrix: Ring Topology}
\label{scat}

The transport through the device is described by the
Landauer--B\"uttiker formula
\ba
I_s &=& \sum_{s'} G_{s s'} V_{s'} + \sum_{t} G_{s t} V_{t} \ , \ s = 1,
\ldots, S \ , \nonumber \\
I_t &=& \sum_{t'} G_{t t'} V_{t'} + \sum_{s} G_{t s} V_{s} \ , \ t = 1,
\ldots, T \ . 
\label{1}
\ea
Here $I_s (I_t)$ is the current through lead $s$ (lead $t$),
respectively, and $V_s (V_t)$ is the voltage applied to that lead. The
conductance coefficients $G_{s s'}$ are given by
\be
G_{s s'} = \sum_{a = 1}^{N_s} \sum_{a' = 1}^{N_{s'}} [ | S_{sa;s'a'}(E,
\Phi)|^2 - \delta_{s s'} ]
\label{2}
\ee
and correspondingly for the index combinations $(s t), (t s)$ and $(t
t')$. The symbol $S_{sa;s'a'}(E, \Phi)$ denotes the element of the
scattering matrix $S(E, \Phi)$ which connects channel $a$ in lead $s$
with channel $a'$ in lead $s'$ at energy $E$ and magnetic flux $\Phi$.
Time--reversal symmetry requires the scattering matrix to obey the
relation
\be
S(E, \Phi) = S^T(E, -\Phi)
\label{3}
\ee
where $T$ denotes the transpose.

The observable ${\cal O}$ describing a given experimental setup is
determined by the experimental arrangement chosen (which of the leads
are grounded, and in which of the leads is a current measured). For any
such setup, ${\cal O}$ will be given as a rational function of the
conductance coefficients $G$. This follows directly from Eq.~(\ref{1}).
In order to present a general framework useful for the analysis of any
such experiment, we focus attention on the scattering matrix $S$. With
the help of the formulas for $S$ given below, it is possible to work
out the dependence of the $G$'s and, hence, of ${\cal O}$ on the
energies of the two Coulomb--blockade resonances, and on the AB phase
$\Phi$.

It is possible to derive the form of $S$ from the Hamiltonian $H$.
This can be done along the lines of Refs.~\cite{hack, maha}. We do not
follow this course here because the explicit solution involves some
lengthy algebra. Rather, we simply present the result which we believe
to be intuitively obvious. In this and the next Section, we focus
attention on the ring geometry and put $V_{\rm L R} = 0$.

The scattering matris $S$ can be written as the product of three
unitary matrices,
\be
S(E, \Phi) = U S^{(\rm res)}(E, \Phi) U^T \ . 
\label{4}
\ee
Without any coupling between each of the QDs and parts 1 and 2 of the
AB device (this condition can be met experimentally by increasing the
heights of the two barriers defining each QD), the resonant part
$S^{(\rm res)}(E, \Phi)$ is equal to the unit matrix, and $S(E, \Phi)$
is, thus, equal to $U U^T$. The form of the latter matrix follows from
the observation that parts 1 and 2 are unlinked. A unitary scattering
matrix $S^{(1)}$ ($S^{(2)}$) describes the non--resonant electron
transport through unlinked part 1 (unlinked part 2, respectively). We
assume that the energy dependence of both matrices is smooth over the
energy interval defined by the widths of the two Coulomb--blockade
resonances introduced below. We accordingly neglect the energy
dependence of both $S^{(1)}$ and $S^{(2)}$. Moreover, both matrices do
not dependent on the magnetic flux $\Phi$, see the remark at the end
of Section~\ref{int}. Time--reversal invariance then implies that both
$S^{(1)}$ and $S^{(2)}$ are symmetric. Thus, we can write for $i =
1,2$ 
\be
S^{(i)} = U^{(i)} [ U^{(i)} ]^T \ .
\label{5}
\ee
Eq.~(\ref{5}) holds for every unitary and symmetric matrix. As
explained in Refs.~\cite{nish,weid}, the unitary transformation
$U^{(i)}$ accomplishes the transformation from the space of physical
channels to the space of eigenchannels. We accordingly write the
matrices $U^{(i)}$ explicitly in the form $U^{(1)}_{s a; \alpha}$ and
$U^{(2)}_{t a; \beta}$. Here $U^{(1)}_{s a;\alpha}$ is the product of
an orthogonal matrix $O^{(1)}_{s a;\alpha}$ which diagonalizes the
symmetric matrix $S^{(1)}$ and of a diagonal matrix with entries
$\exp(i \delta^{(1)}_{\alpha})$ where the $\delta^{(1)}_{\alpha}$'s
are the eigenphaseshifts of $S^{(1)}$, and similarly for $S^{(2)}$.
The index $\alpha$ ($\beta$) runs from 1 to $N_1$ (to $N_2$,
respectively). Here the total number of channels $N_1$ in part 1
($N_2$ in part 2) is given by $N_1 = \sum_{s} N_s$ (by $N_2 = \sum_{t}
N_t$, respectively). The matrix $U$ is defined in the total space of
$N = N_1 + N_2$ channels. It is block--diagonal and given by
\ba
U = \left( \matrix{ U^{(1)} & 0       \cr
                    0       & U^{(2)} \cr} \right) \ .
\label{6}
\ea
Inspection of Eq.~(\ref{4}) shows that for $S^{(\rm res)} = {\bf 1}_N$,
the unit matrix in $N$ dimensions, the scattering matrix $S$ is
block--diagonal and consists of the two matrices $S^{(1)}$ and $S^{(2)}$,
as it must.

It is now obvious that $S^{(\rm res)}$ differs from the unit matrix by
terms which represent the two Coulomb--blockade resonances, one each in
QDL and QDR. Moreover, it is also clear that $S^{(\rm res)}$ is defined
in the space of eigenchannels of both $S^{(1)}$ and $S^{(2)}$. In this
space, the coupling matrix elements $W_{\rho {\rm P}}$ describing the
hopping of an electron from the resonance in QDP (with P = L or R) to
the eigenchannel $\rho$ (with $\rho = \alpha$ for part 1 and $\rho =
\beta$ for part 2) can be shown~\cite{nish} to be real, save for the
AB phase. We accordingly have for $\rho = 1, \ldots, N$ and P = L,R
\be
W_{{\rm P} \rho} = W_{{\rm P} \rho}^* = W_{\rho {\rm P}} \ {\rm unless}
\ {\rm P = L} \ {\rm and} \ \rho = \alpha
\label{7}
\ee
while
\be
W_{{\rm L} \alpha} \exp (i \phi) = W_{\alpha {\rm L}} \exp (- i \phi)
= w_{{\rm L} \alpha}
\label{8}
\ee
with $w_{{\rm L} \alpha}$ real. We note that the $W_{{\rm P} \rho}$'s
differ from but are linear in the $V_{{\rm P}; s a}$'s and $V_{{\rm
    P}; t a}$'s introduced in Section~\ref{hami}.

We can now express $S^{(\rm res)}$ in terms of the matrix elements
$W_{{\rm P} \rho}$, and of the energies $E_{\rm P}$ of the two
Coulomb--blockade resonances. The latter can be varied experimentally
by changing the plunger gate voltage on either QD. We observe that the
matrices $W_{{\rm P} \rho}$ map the space of $N$ eigenchannels onto
the space of the two Coulomb--blockade resonances, and vice versa for
$W_{\rho {\rm P}}$. The matrix $S^{(\rm res)}$ takes the form
\be
S^{(\rm res)}_{\rho \rho'} = \delta_{\rho \rho'} - 2 i \pi \sum_{{\rm P
P'}} W_{\rho {\rm P}} [D^{-1}]_{{\rm P P'}} W_{{\rm {\rm P'}} \rho'} \ .
\label{9}
\ee
The two--by--two matrix $D_{{\rm P P'}}$ has the form (${\rm P} = {\rm
  L,R}$)
\be
D_{{\rm P P'}} = \delta_{{\rm P P'}} [ E - E_{\rm P} ] + i \pi
\sum_{\rho} W_{{\rm P} \rho} W_{\rho {\rm P'}} \ .
\label{10}
\ee
Eqs.~(\ref{2},\ref{4},\ref{9},\ref{10}) constitute the central result
of this Section. It is easy to check that the scattering matrix defined
by these equations is unitary and obeys Eq.~(\ref{3}).

For the benefit of the reader, we rewrite the $S$--matrix in a form
which displays more clearly the physical role of the matrices
$U^{(i)}$ with $i = 1,2$. We define the complex coupling matrix
elements
\ba
\overline{W}_{s a;{\rm P}} &=& \sum_{\alpha} U^{(1)}_{s a; \alpha}
W_{\alpha {\rm P}} \ , \nonumber \\
\overline{W}_{t a;{\rm P}} &=& \sum_{\beta} U^{(2)}_{t a; \beta}
W_{\beta {\rm P}} \ .
\label{10a}
\ea
Then $S$ takes the form
\ba
S_{s a; s' a'} &=& S^{(1)}_{s a; s' a'} - 2 i \pi \sum_{\rm P P'}
\overline{W}_{s a;{\rm P}} [D^{-1}]_{{\rm P P'}} \overline{W}_{{\rm
    P}; s' a'} \ , \nonumber \\
S_{t a; t' a'} &=& S^{(2)}_{t a; t' a'} - 2 i \pi \sum_{\rm P P'}
\overline{W}_{t a;{\rm P}} [D^{-1}]_{{\rm P P'}} \overline{W}_{{\rm
    P}; t' a'} \ , \nonumber \\
S_{s a; t a'} &=& - 2 i \pi \sum_{\rm P P'} \overline{W}_{s a;{\rm
    P}} [D^{-1}]_{{\rm P P'}} \overline{W}_{{\rm P}; t a'} \ ,
\nonumber \\
S_{t a; s a'} &=& - 2 i \pi \sum_{\rm P P'} \overline{W}_{t a;{\rm P}}
[D^{-1}]_{{\rm P P'}} \overline{W}_{{\rm P}; s a'} \ .
\label{10b}
\ea
The matrix $D$ has the same form as in Eq.~(\ref{10}) but can also be
written as
\be
D_{\rm P P'} = \delta_{\rm P P'} [ E - E_{\rm P} ] + i \pi \sum_{s
  a} \overline{W}_{{\rm P}; s a} \overline{W}_{s a ;{\rm P'}} + i \pi
\sum_{t a} \overline{W}_{{\rm P};t a} \overline{W}_{t a ;{\rm P'}} \ . 
\label{10c}
\ee
The transformation~(\ref{10a}) introduces complex matrix elements
$\overline{W}$  which guarantee unitarity of $S$ in the presence of
the non--diagonal unitary matrices $S^{(1)}$ and $S^{(2)}$. 

\section{Analysis. Crossing of Two Resonances}
\label{cross}

The effect of the two resonances which dominate the scattering matrix
is contained entirely in the matrix $D$ defined in Eq.~(\ref{10}). It
is useful to display $D$ in matrix form,
\be
D = \left( \matrix{
E - E_{\rm L} + (i/2) \Gamma_{\rm L} & (i/2) \Gamma_{\rm L R} \cr
(i/2) \Gamma_{\rm R L} & E - E_{\rm R} + (i/2) \Gamma_{\rm R} \cr} \right)
\ ,
\label{11}
\ee
where
\ba
\Gamma_{\rm L} &=& 2 \pi \sum_{\rho} W_{{\rm L} \rho} W_{\rho {\rm L}}
\ , \nonumber \\
\Gamma_{\rm R} &=& 2 \pi \sum_{\rho} W_{{\rm R} \rho} W_{\rho {\rm R}}
\ , \nonumber \\
\Gamma_{\rm L R} &=& 2 \pi \sum_{\rho} W_{{\rm L} \rho} W_{\rho {\rm R}}
\ , \nonumber \\
\Gamma_{\rm R L} &=& 2 \pi \sum_{\rho} W_{{\rm R} \rho} W_{\rho {\rm L}}
\ .
\label{12}
\ea
Our explicit notation combined with Eqs.~(\ref{7}) and (\ref{8}) shows
that $\Gamma_{\rm L}$ and $\Gamma_{\rm R}$ are real, positive and
independent of the magnetic flux $\Phi$, and that the only dependence
on $\Phi$ occurs in $\Gamma_{\rm L R}$ and in $\Gamma_{\rm R L}$. The
latter two quantities are complex and related by
\be
\Gamma_{\rm L R} = \Gamma_{\rm R L}^* \ . 
\label{13}
\ee

We use Eqs.~(\ref{11}) to (\ref{13}) to display the structure of certain
elements of the scattering matrix $S$. We recall that $S$ decays into
two independent scattering matrices $S^{(1)}$ and $S^{(2)}$ whenever we
have $W_{{\rm P} \rho} = 0$ for all P, $\rho$. Parts 1 and 2 of the AB
interferometer are linked only by the two Coulomb--blockade resonances
with energies $E_{\rm L}$ and $E_{\rm R}$. If the two resonance energies
are sufficiently different so that
\be
|E_{\rm L} - E_{\rm R}| \gg \Gamma_{\rm L}, \Gamma_{\rm R},
|\Gamma_{\rm L R}| \ ,
\label{15}
\ee
we can use perturbation theory in $\Gamma_{\rm L R}$ to invert $D$.
Keeping only the lowest--order terms in the expansion, we find
\ba
S_{t b; s a} &=& - 2 i \pi \sum_{\beta \alpha} U^{(2)}_{t b; \beta}
\biggl( W_{\beta {\rm L}} [E - E_{\rm L} + (i/2) \Gamma_{\rm L} ]^{-1}
W_{{\rm L} \alpha} \nonumber \\
&&\qquad + W_{\beta {\rm R}} [E - E_{\rm R} + (i/2) \Gamma_{\rm R}
]^{-1} W_{{\rm R} \alpha} \biggr) U^{(1)}_{s a; \alpha} \ . 
\label{14}
\ea
The two amplitudes on the right--hand side of Eq.~(\ref{14}) can be
interpreted in terms of two paths of the electron on its way from part
1 to part 2. The electron may pass either through QDL (first term) or
QDR (second term). As it passes through QDL, it picks up the AB phase
contained in $W_{{\rm L} \alpha}$. This phase will affect the
interference pattern due to the product of the amplitudes corresponding
to the two paths. Whenever inequality~(\ref{15}) holds, the electron
will not complete one or several loops within the AB ring as it passes
from part 1 to part 2. It is instructive to consider also the terms of
next order. These terms are given by
\ba
&&-2 i \pi \sum_{\beta \alpha} U^{(2)}_{t b; \beta} \biggl( W_{\beta
{\rm L}} [E - E_{\rm L} + (i/2) \Gamma_{\rm L} ]^{-1} \Gamma_{\rm L R}
[E - E_{\rm R} + (i/2) \Gamma_{\rm R} ]^{-1} W_{{\rm R} \alpha} \nonumber
\\ 
&&\qquad + W_{\beta {\rm R}} [E - E_{\rm R} + (i/2) \Gamma_{\rm R} ]^{-1}
\Gamma_{\rm R L} [E - E_{\rm L} + (i/2) \Gamma_{\rm L} ]^{-1} W_{{\rm L}
\alpha} \biggr) U^{(1)}_{s a; \alpha} \ . \nonumber \\
\label{16}
\ea
The path associated with the first amplitude leads the electron first
through QDR and then through QDL, and vice versa for the second amplitude.
With $\Gamma_{\rm L R}$ given by Eq.~(\ref{12}), we see that along the
first path QDL can be reached from QDR either via part 1 or via part 2,
and correspondingly for path 2. In the first (second) case, the AB phase
does not (does) contribute to the scattering amplitude. This is correct
because only in the second case does the electron complete a loop around
the AB ring. A similar analysis of $S_{s a; s' a'}$ and of $S_{t b; t'
b'}$ shows that whenever inequality~(\ref{15}) holds, the scattering is
dominated by $S^{(1)}$ and $S^{(2)}$, respectively. The consecutive
passage through both Coulomb--blockade resonances is strongly inhibited.
We conclude that inequality~(\ref{15}) defines a fairly uninteresting
regime of parameters of the problem.

Interest, therefore, focusses on the regime where this inequality does
not hold and where our perturbation expansion is not appropriate. This
is the regime where the two Coulomb blockade resonances may cross. We
shall see that the crossing displays novel features. Prior to
calculating the exact result, it is useful to visualize the outcome in
terms of a perturbation expansion in powers of $\Gamma_{\rm L R}$ and
$\Gamma_{\rm R L}$. This expansion generates terms of the same form as
in formula~(\ref{16}) but of higher order in $\Gamma_{\rm L R}$ and
$\Gamma_{\rm R L}$. Each propagator $[E - E_{\rm P} + (i/2)
\Gamma_{\rm P} ]^{-1}$ occurring in the expansion signals a visit of
the associated path to QDP with P = L, R. The intermittent factors
$\Gamma_{\rm L R}$ and $\Gamma_{\rm R L}$ signal passage of the
electron from QDR to QDL and vice versa. The passage may proceed via
part 1 or part 2. Thus, the perturbation series stands for the
infinite number of possibilities to connect the channels defined by
the indices of the scattering matrix, by paths. These paths may loop
around the AB ring a number of times, then change direction, loop
again, change direction back etc. until the electron leaves the AB
ring. The AB phase picked up by the electron is the sum of all such
phases picked up in the individual loops and given in terms of the
total number of completed counter--clockwise loops minus the total
number of completed clockwise loops. We are about to calculate the
form of the scattering matrix by diagonalizing the matrix $D$. This
procedure amounts to summing over all the paths just mentioned. This
is why the AB phase will show up in the denominator of the result, see
Eq.~(\ref{20}). (Experimentally, decoherence will actually limit the
number of loops that contribute significantly to the amplitude, see
Ref.~\cite{enss}).

We simplify the algebra by considering an AB ring which contains two
perfectly identical QDs and which itself is perfectly symmetric about
a vertical axis through the middle of Figure~1. Then, $\Gamma_{\rm L}
= \Gamma_{\rm R} = \Gamma$ (this defines the width $\Gamma$). We write
the complex eigenvalues of the matrix $D$ in the form $E -
\varepsilon_i$ with $i = 1,2$. Then
\be
\varepsilon_{1,2} = \frac{1}{2} (E_{\rm L} + E_{\rm R} - i \Gamma) \pm
\frac{1}{2} \sqrt{(E_{\rm L} - E_{\rm R})^2 - |\Gamma_{\rm L R}|^2} \
.
\label{17}
\ee
Let us suppose that we change the resonance energies of both dots in
such a way that $(E_{\rm L} + E_{\rm R})$ is kept fixed while $u =
|E_{\rm L} - E_{\rm R}|$ decreases monotonically from an initially
large value (in the sense of the inequality~(\ref{15})). Then, the
difference $|\varepsilon_{1} - \varepsilon_{2}|$ also decreases
monotonically. Both resonances approach each other, retaining equal
widths. The difference of resonance energies vanishes when $(E_{\rm L}
- E_{\rm R})^2 = |\Gamma_{\rm L R}|^2$: The two resonances coincide in
energy and width. We deal with an exceptional point in the sense of
Ref.~\cite{heis}. At this point, the system possesses only a single
eigenfunction. As we decrease $u$ further, the two resonances
separate, retain equal resonance energies but acquire different
widths. At $u = 0$, the widths differ by $|\Gamma_{\rm L R}|$, the
maximum amount possible.

The value of $|\Gamma_{\rm L R}|$ determines both, the value of $u$
where the resonances coincide and the maximum difference of their
widths. This value depends upon the AB phase $\Phi$. Indeed, from
Eq.~(\ref{12}) we have
\ba
|\Gamma_{\rm L R}|^2 &=& 4 \pi^2 [ ( \sum_{\alpha} w_{{\rm L} \alpha}
w_{\alpha {\rm R}})^2 + ( \sum_{\beta} W_{{\rm L} \beta} W_{\beta {\rm
R}})^2 \nonumber \\
&&\qquad + 2 \cos \phi \sum_{\alpha} w_{{\rm L} \alpha} w_{\alpha {\rm
R}} \sum_{\beta} W_{{\rm L} \beta} W_{\beta {\rm R}}] \ .
\label{18}
\ea
The value of $|\Gamma_{\rm L R}|^2$ oscillates periodically with
magnetic flux $\Phi$ between the maximum value $4 \pi^2 (\sum_{\alpha}
w_{{\rm L} \alpha} w_{\alpha {\rm R}} + \sum_{\beta} W_{{\rm L} \beta}
W_{\beta {\rm R}})^2$ and the minimum value $4 \pi^2 (\sum_{\alpha}
w_{{\rm L} \alpha} w_{\alpha {\rm R}}$ $- \sum_{\beta} W_{{\rm L}
  \beta} W_{\beta {\rm R}})^2$. From Schwarz's inequality we conclude
that the widths of the two resonances are always positive. To estimate
the relative size of $\Gamma$ and of $|\Gamma_{\rm L R}|$, we note
from Eq.~(\ref{12}) that $\Gamma = \Gamma_{\rm L} = \Gamma_{\rm R}$ is
a sum of squares while $\Gamma_{\rm L R}$ is a sum over terms which,
aside from the AB phase, may have either sign. We expect that due to
impurity scattering in parts 1 and 2 of the AB device, the $W_{{\rm P}
  \alpha}$'s are Gaussian random variables, see Ref.~\cite{weid}. As a
consequence, we have $\Gamma \propto N$ while $|\Gamma_{\rm L R}|$
fluctuates strongly with a root--mean--square variance which grows
like $\sqrt{N}$. Thus, the maximum difference of the widths of the two
resonances is expected to be of the order of $\Gamma / \sqrt{N}$.

The possibility of complete coalescence of two resonances displayed
above is a phenomenon which is opposite to the well--known Wigner--von
Neumann level repulsion effect for bound states. The latter occurs
whenever two bound states interact via a Hermitean interaction.
Eq.~(\ref{11}) shows that in the present case, we deal with resonances
with complex resonance energies to begin with, and with a coupling
that is due to a Hermitean interaction multiplied by $i$, the
imaginary unit. This unusual form of interaction occurs because the
two resonances are not coupled directly but via the open channels in
parts 1 and 2. Both differences contribute towards a behavior which
differs from standard Wigner--von Neumann level repulsion. Such
behavior has been discussed previously in the literature. To the best
of our knowledge, the coupling of two resonances was first studied
explicitly by von Brentano {\it et al}.~\cite{bren} in the context of
Nuclear Physics. This work was followed by an experimental
investigation~\cite{micr}. Related work was published in
Ref.~\cite{dohm}. Recent work~\cite{dem} has focussed on the
properties of exceptional points.

To display the features of the exceptional point where $(E_{\rm R} -
E_{\rm L})^2 = |\Gamma_{\rm L R}|^2$ and where $\Gamma_{\rm L} =
\Gamma_{\rm R}$, we consider two slightly asymmetric QDs for which the
two resonance widths $\Gamma_{\rm L}$ and $\Gamma_{\rm R}$ are not
exactly equal. Then, the eigenvalues $\varepsilon_{1,2}$ will never
coincide exactly. This is seen from the expression of the discriminant
which now has the value
\be
\sqrt{(E_{\rm L} - E_{\rm R} - (i/2)(\Gamma_{\rm L} - \Gamma_{\rm
    R}))^2 - |\Gamma_{\rm L R}|^2} \ .
\label{19}
\ee
Imagine now a change of the the parameters of the system in such a way
that the argument of the square root describes a closed loop in the
complex plane around the exceptional point. This could be achieved as
follows. We put $E_{\rm R} = - E_{\rm L} = |\Gamma_{\rm L R}| +
\alpha/2, \Gamma_{\rm L} = - \Gamma_{\rm R} = \beta$ with $\alpha,
\beta$ real and $|\alpha| \ll |\Gamma_{\rm L R}|, |\beta| \ll
|\Gamma_{\rm L R}|$. The discriminant becomes approximately equal to
$\sqrt{2 |\Gamma_{\rm L R}| (\alpha - i \beta)}$. Changing $\alpha$
from a small negative to a small positive value while keeping $\beta >
0$ fixed and small, then keeping $\alpha$ fixed and changing $\beta$
from its small positive value to a small negative one, then keeping
$\beta$ fixed and changing $\alpha$ back to its original value and
doing, finally, the same for $\beta$ yields a rectangle in the complex
plane with the exceptional point in its interior. While under this
operation the phase of $(\alpha - i \beta)$ changes by $2 \pi$, the
phase of $\sqrt{2 |\Gamma_{\rm L R}| (\alpha - i \beta)}$ changes only
by $\pi$: Under this operation, the two eigenvalues
$\varepsilon_{1,2}$ are interchanged, and so are the two
eigenfunctions, including an additional phase factor~\cite{dem}. In
comparison with the work of Refs.~\cite{dem}, the present system seems
to offer an additional degree of freedom in terms of the AB phase. The
latter determines the value of $|\Gamma_{\rm L R}|$, the location of
the exceptional point, and the form of the two eigenfunctions as
linear combinations of the two QD states.

Unfortunately, all these appealing features have no bearing on the
properties of the scattering matrix $S$. This is because the energy
$E$ is always real. As a consequence, we can never reach the
exceptional point, and the determinant of the matrix $D$ never
vanishes for real values of $E$. The two eigenfunctions of $D$ remain
distinct. In view of the recent interest in exceptional points, we
have nevertheless felt that a discussion of this topic is appropriate
in the present context. 

We return to the symmetric case. The matrix $D$ can be diagonalized by
a matrix $A$ so that $D = A^{-1} (E {\bf 1}_2 - \varepsilon) A$ where
$\varepsilon$ denotes the diagonal matrix ${\rm diag}{(\varepsilon_1,
\varepsilon_2)}$. Using this form in Eq.~(\ref{9}), we obtain
\be
S^{(\rm res)}_{\rho \rho'} = \delta_{\rho \rho'} - 2 i \pi \sum_{{\rm
    P P_1 P'}} W_{\rho {\rm P}} A^{-1}_{P P_1} [E -
\varepsilon_{P_1}]^{-1} A_{P_1 P'} W_{{\rm {\rm P'}} \rho'} \ .
\label{20}
\ee
Inserting this matrix into Eqs.~(\ref{2}) yields the conductance
coefficients and, hence, the dependence of any observable on the AB
phase. The AB phase appears explicitly not only in the eigenvalues
$\varepsilon_{1,2}$ but also in the matrix $A$ and, of course, in some
of the $W_{\rho {\rm P}}$'s. The matrix $A$ can easily be calculated.
Details are not given here.

For the sake of completeness, we discuss the limitations of a
two--lead experiment. These limitations have played a role in previous
studies of AB devices~\cite{hack1}. We recall that the scattering
matrix $S$ is unitary and obeys $S^T(E, -\Phi) = S(E, \Phi)$. It follows
that in general, we have $S_{s a; s' a'}(E, -\Phi) = S_{s' a'; s a}(E,
\Phi)$ and correspondingly for the lead indices $(s,t)$ and $(t,t')$.
The cases where the two lead indices coincide are special and yield
$S_{s a; s a'}(E, -\Phi)$ $= S_{s a'; s a}(E, \Phi)$ and $S_{t a; t
a'}(E, -\Phi)$ $= S_{t a'; t a}(E, \Phi)$. For the conductance
coefficients, this means that $G_{s s'}(E, -\Phi)$ $= G_{s' s}(E, \Phi)$,
$G_{t t'}(E, -\Phi) = G_{t' t}(E, \Phi)$ and $G_{s t}(E, -\Phi) = G_{t
  s}(E, \Phi)$ while $G_{s s}(E, -\Phi) = G_{s s}(E, \Phi)$ and $G_{t
  t}(E, -\Phi)$ $= G_{t t}(E, \Phi)$. Unitarity then shows that for a
two--lead experiment the $G$'s are even in $\Phi$ while this is not
the case for the off--diagonal $G$'s when we deal with more than two
leads. This conclusion, first drawn by B\"uttiker~\cite{buet}, is seen
to be quite general and not affected by the topology of our AB ring
with two QDs.

\section{Figure--Eight Topology}
\label{link}

If the two QDs are connected by a wire, the topology differs from that
of a ring analysed so far. The neccessary modifications are quite
straightforward, however. The AB ring is divided into two parts by the
wire connecting the two QDs. Let $\Phi_1 (\Phi_2)$ be the flux through
the lower (the upper) part, respectively. In a manner completely
analogous to Eqs.~(\ref{7}) and (\ref{8}), we put $\Phi_1 (\Phi_2)$
onto $W_{\alpha {\rm L}}$ ($W_{\beta {\rm L}}$, respectively). With
$\phi_i = 2 \pi \Phi_i / \Phi_0$, $i = 1,2$, Eqs.~(\ref{7}) and
(\ref{8}) are thus replaced by
\be
W_{{\rm R} \rho} = W_{{\rm R} \rho}^* = W_{\rho {\rm R}}
\label{21}
\ee
while
\ba
W_{{\rm L} \alpha} \exp (i \phi_1) &=& W_{\alpha {\rm L}} \exp (- i
\phi_1) = w_{{\rm L} \alpha} \ , \nonumber \\
W_{{\rm L} \beta} \exp (- i \phi_2) &=& W_{\beta {\rm L}} \exp (i
\phi_2) = w_{{\rm L} \beta} \ , 
\label{22}
\ea
with $w_{{\rm L} \rho}$ real. A further modification accounts for the
presence of the wire which furnishes a direct link between the two
QDs. We represent this link by a real hopping matrix element $V_{\rm L
  R} = V_{\rm R L}$. This element appears in the matrix $D$ which now
takes the form
\be
D_{{\rm P P'}} = \delta_{{\rm P P'}} [ E - E_{\rm P} ] + i \pi
\sum_{\rho} W_{{\rm P} \rho} W_{\rho {\rm P'}} + (1 - \delta_{\rm P
  P'}) V_{\rm R L} \ . 
\label{23}
\ee
Except for these modifications, all formulas in Section~\ref{scat}
remain unchanged.

For a discussion of the form of the matrix $D$ in Eq.~(\ref{23}), we
distinguish two limiting cases, where $|\Gamma_{\rm L R}|$ dominates
$|V_{\rm R L}|$ or vice versa. It is obvious that for $|V_{\rm R L}|
\ll |\Gamma_{\rm L R}|$ we (approximately) retrieve our previous
results since $\phi_1 + \phi_2 = \phi$. The distribution of the AB
phase over two sets of matrix elements only complicates the notation.
Therefore, the interesting novel limiting case is the one where
$|V_{\rm R L}| \gg |\Gamma_{\rm L R}|$. We neglect $\Gamma_{\rm L R}$
in comparison with $V_{\rm R L}$ and consider again the symmetric case
with $\Gamma_{\rm L} = \Gamma_{\rm R} = \Gamma$. Explicitly, the
matrix $D$ is given by
\be
D = \left( \matrix{
E - E_{\rm L} + (i/2) \Gamma & V_{\rm L R} \cr
V_{\rm R L} & E - E_{\rm R} + (i/2) \Gamma \cr} \right) \ .
\label{24}
\ee
The matrix $D$ does not depend upon the AB phase (which now appears
only in the matrix elements $W$ in Eq.~(\ref{9})). Moreover, the
interaction $V_{\rm L R}$ causes standard level repulsion between the
two resonances. The AB phase dependence of the conductance
coefficients becomes complicated not because of the matrix $D$ but
because the electron may traverse several different paths on its way
from the entrance channel to the exit channel. For instance, if the
source (sink) is located in part 1 (part 2) of the AB device, there
are four possible paths. One enters and leaves QDL, one enters and
leaves QDR, one enters QDL but leaves QDR, and one enters QDR and
leaves QDL. The relative weight of the four contributions depends upon
the eigenvectors and eigenvalues of the matrix $D$. Again, these can
be worked out straightforwardly.

Corrections to these limiting cases can easily be calculated in terms
of a power--series expansion in $V_{\rm L R}$, or in $\Gamma_{\rm L R}$
and $\Gamma_{\rm R L}$. A full diagonalization of the matrix $D$ in
Eq.~(\ref{23}) is also possible, covers all the intermediary cases, and
yields interesting results. The eigenvalues $\varepsilon_{1,2}$ are
given by
\ba
&&\varepsilon_{1,2} = \frac{1}{2} (E_{\rm L} + E_{\rm R} -
(i/2)(\Gamma_{\rm L} + \Gamma_{\rm R})) \nonumber \\
&&\pm \frac{1}{2} \sqrt{[E_{\rm L} - E_{\rm R} - (i/2)(\Gamma_{\rm L}
  - \Gamma_{\rm R})]^2 + 4 [V_{\rm L R} + (i/2) \Gamma_{\rm L R}] [
  V_{\rm R L} + (i/2) \Gamma_{\rm R L}]} \ . \nonumber \\
\label{25}
\ea
The eigenvalues coincide whenever the argument of the square root
vanishes, i.e., whenever
\be
[E_{\rm L} - E_{\rm R} - (i/2)(\Gamma_{\rm L} - \Gamma_{\rm R})]^2 = -
4 [ V_{\rm L R} + (i/2) \Gamma_{\rm L R}] [ V_{\rm R L} + (i/2)
\Gamma_{\rm R L} \ .
\label{26}
\ee
Eq.~(\ref{26}) extends the definition of an exceptional point to the
figure eight topology. We note that the right--hand side of
Eq.~(\ref{26}) is a periodic function of $\phi$.

\section{Summary and Discussion}
\label{summ}

We have presented a very general approach to the transport properties
of an AB device containing two QDs. Our main assumptions are:

(i) The electrons do not interact. Then, we can use the
Landauer--B\"uttiker approach and express every observable in terms of
the conductance coefficients $G$. The latter are given as squares of
the elements of the scattering matrix $S$.

(ii) For the description of the the two Coulomb--blockade resonances,
we use the single--level approximation.

(iii) The only relevant energy dependence of $S$ is due to the two
Coulomb--blockade resonances, one in either QD. Then, scattering in
parts 1 and 2 of the AB device is independent of energy, and the
scattering matrix $S$ attains the form of Eq.~(\ref{4}), with
$U^{(1)}$ and $U^{(2)}$ independent of energy and AB phase.

Under these assumptions, we have presented a comprehensive description
of an AB device with the topology of a ring or of a figure eight. In
particular, we have displayed explicitly the dependence of the
$S$--matrix upon energy and AB phase. We have shown that a novel
situation arises in the case of a ring topology. Here the two
resonances (with complex energies) are coupled via the channels in
part 1 and part 2 of the AB device. This coupling is given by a
Hermitean matrix multiplied by the imaginary unit $i$. This case
differs fundamentally from the standard coupling of two bound states
by a Hermitean interaction. The latter case leads to level repulsion,
the former may lead to coalescence of levels. It seems that this
phenomenon has been observed in Ref.~\cite{enss}.

We now address the approximations we have made. Perhaps most
importantly, we have neglected the Coulomb interaction between the
two electrons populating the two QDs, and that between each of these
and the electrons on either QD. Inclusion of the Coulomb interaction
would make it impossible to use the Landauer--B\"uttiker approach as
we have done. Alternatives are discussed in a recent
review~\cite{alei}. The standard procedure employs rate equations for
the occupation probabilities of the single--particle levels. However,
this approach is manifestly unsuited to deal with phase correlations
between scattering amplitudes. The latter are of central importance
for an AB device. A more elaborate approach~\cite{alei} uses a
description in terms of an effective Hamiltonian. This approach
assumes that the single--particle states in the QD are described by
random--matrix theory. The effective Hamiltonian for the isolated QD
is obtained as the leading term in a systematic expansion in inverse
powers of $g$, the dimensionless conductance. The two--terminal
conductance is then obtained from the Kubo formula and another
effective Hamiltonian which includes the coupling to the leads. The
latter is determined via a non--trivial theoretical derivation which
in turn involves approximations. To the best of our knowledge, this
approach has never been used for a multi--terminal device involving an
AB ring. Therefore, it is not known whether the approach is able to
account for the phases which are relevant for the present system. At
sufficiently low temperatures, the Coulomb interaction leads to
Kondo--like effects in QD's. Remarkably, the calculation of the phase
of a QD embedded in an AB ring has recently been worked out in the
Kondo regime~\cite{silv}, in spite of the difficulties just mentioned
to deal with the Coulomb interaction outside this regime.

In view of this situation, we can only offer a few qualitative remarks
in support of the present approach. First, the Coulomb interaction has
likewise been neglected in Refs.~\cite{weid,hack,hack1} which addressed
the AB phase for a single QD embedded in an AB ring. The results offered
what seems a realistic and useful description of the overall phase
dependence of experimental observables. Second, our use of the
single--level approximation for each QD lends greater plausibility to
the inclusion of the charging energy in the definition of the energies
labelled $E_{\rm P}$. We admit, however, that the Coulomb energy
between the two electrons (one on each QD) is not covered by this
argument. Our neglect of the Coulomb interaction is not restricted to
the neglect of the charging energy. We have likewise neglected the
spin--dependent interaction between electrons. The latter is induced
via the exchange term and lifts the degeneracy between singlet and
triplet states~\cite{burk,alei}. This spin--dependent interaction
plays a prominent role in Kondo--type effects. We expect that this is
likewise the case in the present situation, especially when the two
resonances overlap. Therefore, our approach can only be expected to
work above the Kondo temperature.

We believe that our other approximations are less severe. The
single--level approximation should work at and near an isolated
Coulomb--blockade resonance whenever resonance width and temperature
are small compared to the charging energy. The neglect of all other
energy dependence but that due to the resonances in the scattering
matrix should be excellent barring very special circumstances.

{\bf Acknowledgment.}The author learned of the
experiments~\cite{enss,huet} at a workshop on zero--dimensional
conductors held at the Max Planck Institut f\"ur Physik
komplexer Systeme in November 2002 in Dresden. He is grateful to the
organizers for having invited him. He is also grateful to P. von
Brentano for a discussion and useful suggestions. He thanks K. Ensslin
for a copy of the Diploma thesis by M. Sigrist~\cite{enss}, and both A.
H\"uttel and C. Dembowski for a communication.

\begin{figure}
\center
\includegraphics[width=10cm]
{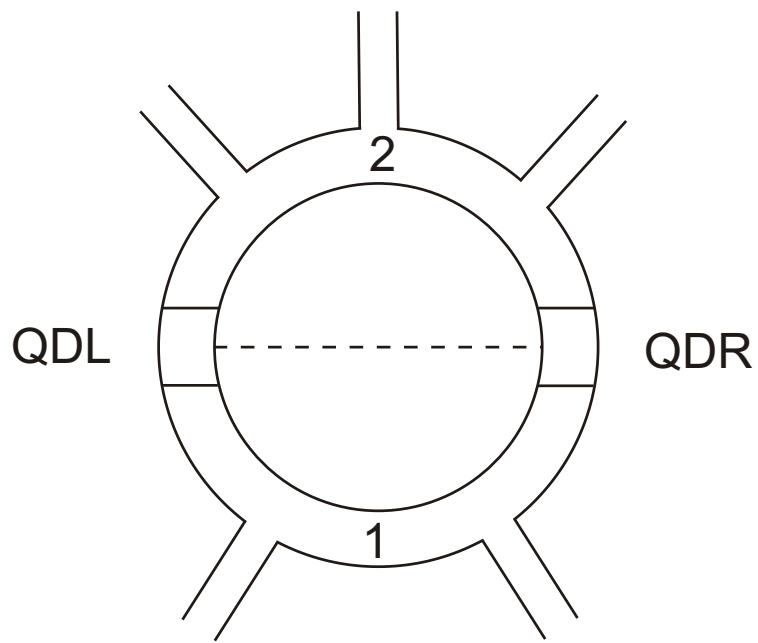}
\caption{Schematic representation of an AB interferometer with 5
  external leads and two quantum dots labelled QDL and QDR embedded
  into its arms. The dotted line represents a link between the two
  quantum dots.}
\label{fig:Figure1}
\end{figure}


\begin{thebibliography}{99}

\bibitem{enss}K. Ensslin, private communication, and M. Sigrist,
Diploma Thesis, ETH Z\"urich (unpublished).

\bibitem{huet}A. H\"uttel, private communication. See also: A. W.
Holleitner et al., Science 297, 70 (2002) and A.W. Holleitner et al.,
Phys. Rev. Lett. 87, 256802 (2001). 

\bibitem{burk}G. Burkard, D. Loss, and D. P. DiVincenzo,
  Phys. Rev. {\bf B 59}, 2070 (1999).

\bibitem{hack}G. Hackenbroich and H. A. Weidenm\"uller, Phys. Rev.
{\bf B 53}, 16379 (1996) and Europhys. Lett. {\bf 38}, 129 (1997).

\bibitem{weid}H. A. Weidenm\"uller, Phys. Rev. {\bf B 65}, 245322
(2002).

\bibitem{hack1}G. Hackenbroich, Phys. Rep. {\bf 343}, 464 (2001).

\bibitem{maha}C. Mahaux and H. A. Weidenm\"uller, Shell--Model
  Approach to Nuclear Reactions, North--Holand Publishing Company,
  Amsterdam (1969).

\bibitem{nish}H. Nishioka and H. A. Weidenm\"uller, Phys. Lett. {\bf
157B}, 101 (1985).

\bibitem{heis}W. D. Heiss, Eur. Phys. J. {\bf D 7}, 1 (1999);
  Phys. Rev. {\bf E 61}, 929 (2000).

\bibitem{bren}P. von Brentano and M. Phillipp, Phys. Lett. {\bf 454B},
   171 (1999). See also P. von Brentano, Phys. Lett. {\bf 238B}, 1
  (1990) and Nucl. Phys. {\bf A 550}, 143 (1992).

\bibitem{micr}M. Phillipp, P. von Brentano, G. Pascovici, and
  A. Richter, Phys. Rev. {\bf E 62}, 1922 (2000).

\bibitem{dohm}H. Estrada, L. S. Cederbaum, and W. Domcke, J. Chem
  Physics {\bf 84}, 1 (1986).

\bibitem{dem}C. Dembowski {\it et al.}, Phys. Rev. Lett. 86, 787
  (2001) and Phys. Rev. Lett. 90, 034101 (2003).

\bibitem{buet}M. B\"uttiker, Phys. Rev. Lett. {\bf 57}, 1761 (1986). 

\bibitem{alei}I. L. Aleiner, P. W. Brouwer, and L. I. Glazman, Phys.
Rep. {\bf 358}, 309 (2002).

\bibitem{silv}P. G. Silvestrov and Y. Imry, cond-mat/0112308.

\end{thebibliography}
\end{document}